# Lateral depletion effect on two-dimensional ordering of bacteriorhodopsins in a lipid bilayer: A theoretical study based on a binary hard-disk model

*Keiju Suda[1], Ayumi Suematsu[2], and Ryo Akiyama[1]\**

[1]*Department of Chemistry, Graduate School of Science, Kyushu University, Fukuoka 819-0395, Japan*

[2]*Faculty of Science and Engineering, Kyushu Sangyo University, Fukuoka 813-8503, Japan*

\* rakiyama@chem.kyushu-univ.jp

The two-dimensional ordering of bacteriorhodopsins in a lipid bilayer was studied using a binary hard-disk model. The phase diagrams were calculated, taking into account the lateral depletion effects. The critical concentrations of the protein ordering for the monomers and the trimers were obtained from the phase diagrams. The critical concentration ratio agreed well with the experiment when the repulsive core interaction between the depletants, namely the lipids, was taken into account. The results suggest that the depletion effect plays an important role in the association behaviors of transmembrane proteins.

The bacteriorhodopsin (bR) is a kind of trans-membrane protein. Wild-type bRs make trimers and they form a two-dimensional (2D) ordering structure in a lipid bilayer [1,2]. The bR rich parts, called purple membranes, work as proton pumps in *Halobacterium salinarum*. Some experiments suggest that the bR ordering stabilizes the functional structure of bR in the photo cycle [3,4]. The stability of the ordering is also affected by the condition of the membrane [5]. However, some kinds of mutant bRs construct the 2D ordering structure, although they do not form trimers [6]. The critical concentrations (CC) of the mutant bRs for 'crystallization' are much larger than those of the wild-type bRs [7]. The ratio of the monomer-CC to the trimer-CC (CCR) is about 10.2 [7].

The driving force for the ordering of the trans-membrane proteins has not yet been clarified. There is no covalent bond between bRs in the membrane [8]. Moreover, stabilizing the 2D structure based on the Coulomb interactions and hydrogen bonds between the protein sur- faces can be expected to be difficult because the hydration of hydrophilic surfaces is strongly stabilized in an aqueous phase rather than in the membrane. Had the effects of ionic bonds and hydrogen bonds been large enough, the proteins would be dissolved in the aqueous phase. By contrast, lateral depletion interactions between the proteins are expected [9,10]. Basically, the lipid molecules and the transmembrane proteins are confined to the lipid bilayer. The pseudo- 2D space is crowded by the lipid molecules and the proteins. Therefore, in the present work, the depletion interaction caused by the lateral translational motion of lipid molecules was studied.

Because the exposure of the bRs and the lipid molecules to an aqueous phase must pay large penalties in free energy, the motions of these molecules are confined to the psudo-2D space. Hence, the 2D binary hard-disk model was adopted (FIG.1). The diameter $\sigma_{tri} = 6.2$ nm for the trimer bR was estimated from an electron microscope image [1,2,11]. The estimated monomer diameter was about 3.0 nm. Three monomer diameters $\sigma_{mono} = 2.9$, 3.0, and 3.1nm were examined to remove the arbitrariness for the model. The small disks were lipid molecules with the diameter $\sigma_{lip} \sim 0.5$ nm [12]. For the same reason, we examined three diameters $\sigma_{lip} = 0.4, 0.5,$ and $0.6$ nm.

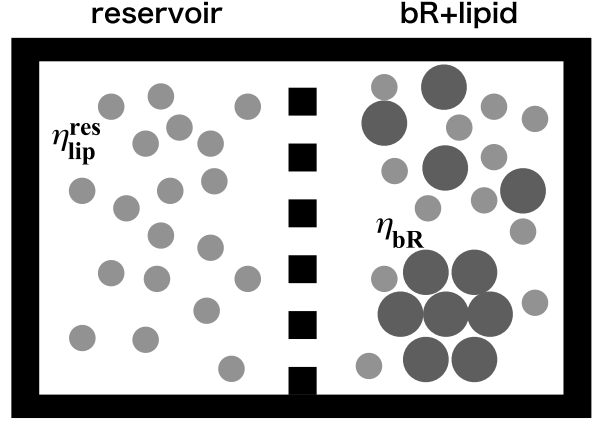

**Fig. 1.** The binary hard-disk system (right side) with a reservoir of small disks (left side). The bRs and the lipid molecules are modeled as larger and smaller disks, respectively. The packing fraction of lipid molecules $\eta_{lip}$ in the binary hard-disk system is controlled by $\eta_{lip}^{res}$ and $\eta_{bR}$.

The isothermal system, consisting of the binary hard disks, contacts with a reservoir that contains only small disks (lipid molecules). The system is in osmotic equilibrium with the reservoir. The depletion effect in the system is controlled by the packing fraction of small disks in the reservoir. The coexistence curves of the bR fluid and the ordered phases were obtained by two equations, as follows:

$$p_f\left(\eta_{bR}^{fluid}, \eta_{lip}^{res}\right) = p_{ord}\left(\eta_{bR}^{ordered}, \eta_{lip}^{res}\right), \qquad (1)$$

$$\mu_f\left(\eta_{bR}^{fluid}, \eta_{lip}^{res}\right) = \mu_{ord}\left(\eta_{bR}^{ordered}, \eta_{lip}^{res}\right), \qquad (2)$$

where $p_f, p_{ord}, \eta_{bR}^{fluid}, \eta_{bR}^{ordered}, \eta_{lip}^{res}, \mu_f, \mu_{ord}$ are the bR pressures for the fluid and ordered phases, the bR packing fractions for the fluid and ordered phases, the packing fraction of lipid for the reservoir, and the bR chemical potentials for the fluid and ordered phases, respectively. Therefore, the values for $p_f\left(\eta_{bR}^{fluid}, \eta_{lip}^{res}\right), p_{ord}\left(\eta_{bR}^{ordered}, \eta_{lip}^{res}\right), \mu_f\left(\eta_{bR}^{fluid}, \eta_{lip}^{res}\right)$, and $\mu_{ord}\left(\eta_{bR}^{ordered}, \eta_{lip}^{res}\right)$ were calculated. Basically, the method was similar to that explained in a textbook [13]. While the textbook describes three-dimensional (3D) systems, the present system is 2D. Not only the 2D Carnahan–Staring equation of state (2D-CSE) [14] was examined, but also the 2D scaled particle theory (2D-SPT) [15,16] for the pure bR fluid phases.

First, the pressure and the chemical potential of the pure bR system for the fluid and the ordered phase were obtained. The pressure of pure bRs for the fluid phase $p_f^0$ is obtained by 2D-CSE as follows:

$$\beta p_f^0 v_{bR} = \frac{\eta_{bR} + \frac{(\eta_{bR})^2}{8}}{(1 - \eta_{bR})^2}, \qquad (3)$$

where $\eta_{bR}, \beta,$ and $v_{bR}$ are the packing fractions of the bRs, $1/k_B T$, and the area for a bR, respectively; $k_B$ is the Boltzmann constant and $T$ is the absolute temperature. The chemical potential of pure bRs for the fluid phase $\mu_f^0$ is obtained as

$$\beta\mu_f^0 = \ln\left[\frac{\Lambda^2}{v_{bR}}\right] + \ln[\eta_{bR}] - \frac{7}{8}\ln[1-\eta_{bR}] + \frac{7}{8(1-\eta_{bR})} + \frac{9}{8(1-\eta_{bR})^2} - 2, \qquad (4)$$

where $\Lambda = h(2\pi m_{bR} k_B T)^{-1/2}$ is the thermal de Broglie wavelength in the 2D space. h and $m_{bR}$ are the Planck constant and the mass for one bR, respectively. By using 2D-SPT, $p_f^0$ and $\mu_f^0$ were also obtained, as follows:

$$\beta p_f^0 v_{bR} = \frac{\eta_{bR}}{(1-\eta_{bR})^2}, \qquad (5)$$

$$\beta\mu_f^0 = \ln\left[\frac{\Lambda^2}{v_{bR}}\right] + \ln\left[\frac{\eta_{bR}}{1-\eta_{bR}}\right] + \frac{2-\eta_{bR}}{(1-\eta_{bR})^2} - 2. \qquad (6)$$

By contrast, the pressure and the chemical potential of the 2D crystal of pure bRs were substituted with the values of the 2D-ordered state. The free volume theory for a 2D system (2D-FVT) [17] was adopted for this calculation. The pressure and chemical potential for the hexagonal lattice are

$$\beta p_{ord}^0 v_{bR} = \frac{2\eta_{bR}}{1-\frac{\eta_{bR}}{\eta_{cp}}}, \qquad (7)$$

$$\beta\mu_{ord}^0 = \ln\left[\frac{\Lambda^2}{v_{bR}}\right] - 2\ln\left[\frac{\eta_{cp}}{\eta_{bR}}-1\right] + \frac{2\eta_{cp}}{(\eta_{cp}-\eta_{bR})}, \qquad (8)$$

where $\eta_{cp} = \pi/2\sqrt{3}$ is the packing fraction at close packing.

The adoption of the hexagonal lattice in the present study is discussed here. The first problem is the possibility of existence of a long-range positional order. Theoretical studies, such as the Mermin–Wagner theorem, seemed to rule out the existence of an infinite long- range positional order [18,19]. By contrast, a 2D hexagonal bR ordering structure has been observed experimentally on the purple membrane of the bacteria [1,2]. The diameter of the ordering structure is large, but is not infinite. It is at most 0.5 μm and the patch has fewer than 6000 bR trimers [20]. It seems that the finite ordering structure is stable in the bacteria. We think that this ordered state can be treated approximately as a crystal.

The second problem is the choice of the solid structure. A hexatic phase appears near the solid phase in one component's hard-disc system [21-24]. It means that the hexatic phase is most stable in the region. The fragility of the hexatic phase was shown in a recent simulation study [25]. According to the study of a binary mixture of hard disks, the hexatic phase shrinks and disappears as small disks are introduced into the system of large disks. The shrinking appears at very low concentrations of the small disk. In the present study, we discuss the behaviors of the binary system and the concentrations of the small disk, i.e., the lipid molecule, which were medium and high. Therefore, the hexagonal lattice was adopted as the ordered state.

We obtained the semi-grand potential $\Omega(N_{bR}, V, T, \mu_{lip})$ by 2D-FVT as follows:

$$\Omega(N_{bR}, V, T, \mu_{lip}) = F^0(N_{bR}, V, T) - p^{res}\langle V_{free}^{mix}\rangle. \tag{9}$$

Here, $F^0$ is the Helmholtz free energy of a bR-pure system, $p^{res}$ is the pressure of the reservoir system, and $\langle V_{free}^{mix}\rangle$ is the free-area of lipid in the lipid–bR system. We substitute the pressure of 2D-SPT for $p^{res}$. The $\langle V_{free}^{mix}\rangle$ is obtained using the scaled particle theory as follows:

$$\langle V_{free}^{mix}\rangle = V^{mix}\alpha. \tag{10}$$

Here, α is defined as

$$\alpha \equiv \exp[-\beta W], \tag{11}$$

$V^{mix}$ is the area of the lipid–bR system, and $W$ is the work to insert a lipid molecule in the bR-pure system. α is calculated using 2D-SPT as follows:

$$\alpha = (1 - \eta_{bR})\exp\left[-\frac{2\eta_{bR}q}{1 - \eta_{bR}} - \frac{q^2\eta_{bR}}{(1 - \eta_{bR})^2}\right]. \tag{12}$$

Here, $q$ is the diameter ratio between a bR (monomer or trimer) and lipid. The pressure and chemical potential are obtained from $\Omega(N_{bR}, V, T, \mu_{lip})$, as follows,

$$p = p^0 + p^{res}\alpha - p^{res}\eta_{bR}\left(\frac{\partial\alpha}{\partial\eta_{bR}}\right)_{N_{bR},T,\mu_{lip}}, \tag{13}$$

$$\mu_{bR} = \mu_{bR}^0 - p^{res}v_{bR}\left(\frac{\partial\alpha}{\partial\eta_{bR}}\right)_{V,T,\mu_{lip}}, \tag{14}$$

$$\left(\frac{\partial\alpha}{\partial\eta_{bR}}\right) = \frac{-\eta_{bR}^2 + (-q^2 + 2q + 2)\eta_{bR} - q^2 - 2q - 1}{(1 - \eta_{bR})^2}\exp\left[-\frac{2\eta_{bR}q}{1 - \eta_{bR}} - \frac{\eta_{bR}q^2}{(1 - \eta_{bR})^2}\right], \tag{15}$$

where $p^0$ and $\mu_{bR}^0$ are pressure and chemical potential in the bR-pure system.

The free energies of pure bR-ordered state and of pure fluid phase were calculated by using 2D-FVT and 2D-CSE, respectively. The phase diagrams for the bR trimers (solid) and bR monomers (dots) are shown in FIG. 2(a). The *y* axis shows the packing fraction of lipid molecules in the reservoir and the *x* axis shows that of bRs. In the case of the pure bR system ($\eta_{lip}^{res} = 0$) the region $0.703 < \eta_{bR} < 0.747$ is the coexistence region (fluid + ordered state).

The coexistence regions of the proteins expand around $\eta_{lip}^{res} = 0.35$. This expansion appears not only at the boundary of the fluid side, but also at the boundary of the ordered state side. The fluid side boundary decreases monotonically and there is no critical point. The coexistence region for the bR trimer ($q = 0.08065$) is wider than that for the bR monomer ($q = 0.16667$). In other words, as the parameter $q$ decreases, the coexistence region becomes wider. This $q$-dependence of the width for the coexistence region is consistent with the $q$- dependence of the effective attraction between large disks. It is for this reason that the ordered state appears at the lower concentration when the effective attraction becomes stronger [9,10] (see the Supplemental Material [26]).

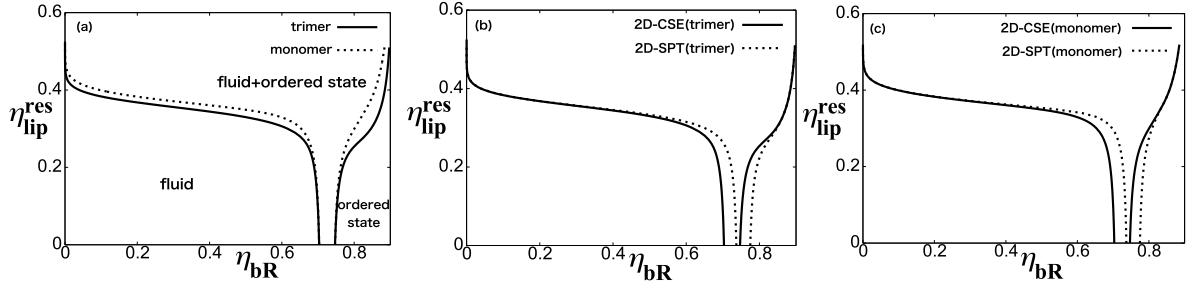

**Fig. 2.** (a) Phase diagrams of the bR trimer (solid curves) and the bR monomer (dotted curves). The free energy of pure bR fluid was calculated on the basis of 2D-CSE. (b) Phase diagrams of the bR trimer. The free energy of pure bR fluid was calculated on the basis of 2D-CSE (solid curves) and 2D-SPT (dotted curves). (c) Phase diagrams of the bR monomer. The free energy of pure bR fluid was calculated on the basis of 2D-CSE (solid curves) and 2D-SPT (dotted curves).

The bR-ordered phase appears at the boundary of the fluid side, and the concentration at the boundary is the CC for the ordered phase. The CC value of the trimers is lower than that of the monomers. For example, when the $\eta_{lip}^{res}$ was 0.4, the CC for the trimers was 0.042 and that for the monomer 0.100. The difference obtained by our calculation qualitatively agrees with that obtained by the experiments [7].

When the $\eta_{lip}^{res}$ is larger than 0.35, the fluid side boundary is shifted to the very low packing fraction $\eta_{bR}$, as mentioned above. This shift is explained based on the depletion effects induced by the lateral translational motion of lipid molecules, as follows. The depletion forces between bRs become stronger with increasing the packing fraction of the lipid molecules. This effective attraction causes the sudden boundary shift around $\eta_{lip}^{res} = 0.35$ [9,10] (see the Supplemental Material [26]).

In this 2D system, the high–low density fluids coexistence region does not exist because of the absence of the critical point. It means that the bR disordered 2D condensed phase is not observed on the membrane. In FIG. 2(a), 2D-CSE was adopted to obtain the free energy for the reference fluid phase; namely, the pure bR fluid phase. Another theory, namely 2D-SPT, was adopted for comparison. The free energy for the reference ordered phase was calculated by the 2D-FVT, again. The theory and connection of the reservoir are also similar to the calculation for FIG. 2(a). The phase diagrams are shown in FIG. 2(b) and (c) to compare with those given by the 2D-CSE and 2D-FVT.

For the pure bR system ($\eta_{lip}^{res} = 0$) the coexistence region calculated by using 2D-SPT was $0.737 < \eta_{bR} < 0.775$. The coexistence region calculated by using 2D-SPT shifted to a higher $\eta_{bR}$ than that calculated by using 2D-CSE. This difference was caused by the difference between 2D-CSE and 2D-SPT, because there was no depletant in this system. However, the depletion effect becomes stronger as the $\eta_{lip}^{res}$ increases.

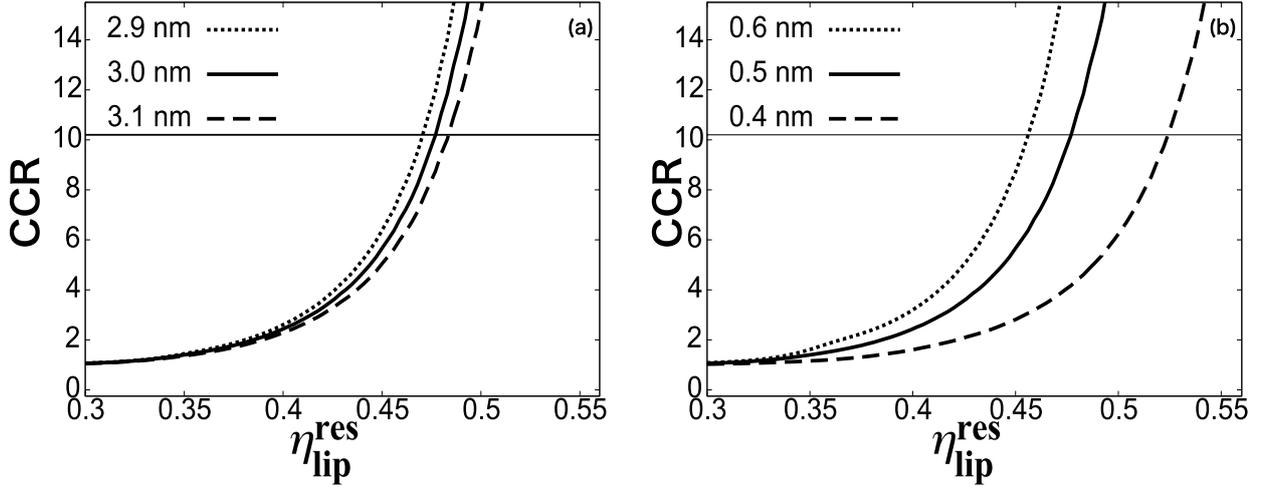

**Fig. 3.** Calculated CCR of bR trimers to monomers. The horizontal thin solid line shows the experimental value, 10.2. (a) The diameter of the lipid molecule is 0.5 nm. The dotted curve (bR monomer diameter: 2.9 nm), the solid curve (bR monomer diameter: 3.0 nm), the dashed curve (bR monomer diameter: 3.1 nm). (b) The diameter of bR monomer is 3.0 nm. The dotted curve (the lipid diameter: 0.6 nm), solid curve (the lipid diameter: 0.5 nm), dashed curve (the lipid diameter: 0.4 nm).

As the $\eta_{lip}^{res}$ increases, the difference between phase diagrams based on 2D-SPT and 2D-CSE disappears. When the $\eta_{lip}^{res} > 0.35$, the solid and the dotted curves overlap in FIG. 2(b) and (c). This overlap indicates that the depletion effects become more dominant for bR ordering than the reference system as the depletion effect induced by lipid molecules increases. That is, when the $\eta_{lip}^{res}$ is above 0.35, the depletion force almost determines the CC for bR ordering. Therefore, it seems that the difference between the theories of the reference system disappears.

Some results for the 2D one-component hard-disk system are brought here from references [21,27] to discuss the present results. According to the simulation study, the boundary between the fluid and the coexistence regions exists at 0.700 [21]. An experimental study gives the value 0.68 for the boundary [27]. The fluid phases disappear at 0.716 (simulation) [21] and 0.70 (experiment) [27]. The fluid and hexatic phases disappear when the packing fraction increases. The monophase for solid becomes most stable under the high packing fraction region. The boundary between the hexatic and solid phases is 0.720 (simulation) and 0.73 (experiment), respectively [21,27].

In the present study, the hexagonal crystal structure was adopted as the ordered phase, and the coexistence regions for the fluid-ordered phase appear 0.703 − −0.747 (for 2D-CSE and 2D-FVT) and 0.737 − −0.775 (for 2D-SPT and 2D-FVT). These values deviate from the exact values given by the simulations and the experiments. However, it does not mean that the present results are meaningless to obtain the CCR. Because of the depletion effect, these differences on the boundaries disappear at the packing fraction of the bio-membrane ($\eta_{lip}^{res} = 0.5$) [12]. FIG. 2(b) and (c) show that the CC boundaries virtually agree with each other in the region $\eta_{lip}^{res} > 0.38$, suggesting that the effective attraction is strong enough. We, therefore, assume that the argument of the hexatic phase for the reference bR system is avoided for the CCR in the present study.

CCRs were obtained for the comparison between the calculated and experimental results. CCR becomes larger than 1 because the CC for the bR trimer is lower than that for the bR monomer. The calculated CCR results are shown in FIG. 3(a) and (b). In all models, the CCRs monotonically increase as the $\eta_{lip}^{res}$ increases. The increase in CCR becomes steeper as the monomer diameter decreases (FIG. 3(a)). By contrast, the increase in CCR becomes steeper as the lipid diameter increases (FIG. 3(b)). Hence, the CCR curves depend on the model. Here, 2.9 and 3.1 nm are small enough and large enough for the monomer size, respectively; 0.4 and 0.6 nm are also too small and too large for the lipid size, respectively. Therefore, the ranges for the parameters are wide enough to discuss the ordering mechanism.

The CCR experimental value is about 10.2 [7]. For example, when the diameters of the lipid molecule and the bR monomer are 0.5 nm and 3.0 nm, respectively, the calculated CCR agrees with the experimental CCR at the $\eta_{lip}^{res} = 0.477$. This result is important. According to literature [12], the lipid number density of a cell membrane monolayer is $2.5 \times 10^6 \#/\mu m^2$. When the monomer diameter is 2.9, 3.0, and 3.1 nm, the calculated CCR agrees with the experimental CCR at the lipid number densities of = 2.40, 2.43, and $2.46 \times 10^6 \#/\mu m^2$, respectively. Each value is almost the same, about $2.4 \times 10^6 \#/\mu m^2$. This value is reasonable compared with $2.5 \times 10^6 \#/\mu m^2$.

When the lipid diameter is 0.4, 0.5, and 0.6 nm, the calculated CCR agrees with the experimental CCR at the lipid number densities of = 4.17, 2.43, and $1.61 \times 10^6 \#/\mu m^2$, respectively. The calculated results are reasonable because the calculated lipid number density of $2.43 \times 10^6 \#/\mu m^2$ is very close to the estimated value of $2.5 \times 10^6 \#/\mu m^2$ for a cell membrane monolayer. The orders of 4.17 and $1.61 \times 10^6 \#/\mu m^2$ are also the same as that of $2.5 \times 10^6 \#/\mu m^2$.

Calculated results suggest that the driving force for the ordering is the depletion effect. Although the model is simple, the calculated CCR is very close to the experimental value. Indeed, the calculated results depend on the size ratios between disks. However, the calculated results remain reasonable if the parameters for the models are chosen in realistic values. Therefore, it seems that this conclusion is robust.

An ideal gas model was also examined as a lipid reservoir. The depletion effect for the ideal gas model became much smaller than for 2D-CSE and 2D-SPT. The wide coexistence region appears when $\eta_{lip}^{res}$ is larger than 0.9 (figure not shown). However, it is much larger than the reasonable value for the present system because the packing fraction of the 2D close-packed structure is 0.907. That is, the ideal gas model does not explain the experimental results quantitatively. This disagreement contrasts with numerically appropriate results for 2D-SPT and 2D-CSE. As described above, the calculated results agree with experimental results when the lipid molecules in the reservoir interact with each other with a repulsive force. The equations of state for 2D-SPT and 2D-CSE show that the pressures are much higher than for the ideal gas because each lipid molecule has a repulsive core, and they are crowded in the reservoir. The equation of state for 2D-SPT in the fluid phase is known as a very precise equation. Thus, the effective attraction between bRs becomes stronger than that for the ideal gas reservoir. Therefore, the results indicate that the crowding in the reservoir is important in the calculation of the depletion force between bRs.

The reduced second virial coefficients were also estimated $B_2/B_2^{HD}$ based on the Asakura-Oosawa (AO) theory [9,10]. In this theory, the depletants do not interact with each other. In other words, the depletants are ideal gas molecules. The effective attraction between large disks becomes weaker than the two-component hard-disk system. The coefficients $B_2/B_2^{HD}$ do not have any negative values in $q = 0.05 - -0.3$, even when $\eta_{lip}^{res} = 0.5$. In the present study, the ratios $q$s were 0.08065 and 0.16667, indicating that the depletion effect of the ideal gas reservoir was too weak to discuss the condensation. By contrast, the coefficients $B_2/B_2^{HD}$ become reasonable negative values if the modified AO (MAO) theory is adopted (see the Supplemental Material [26]). In the MAO theory, the ideal gas pressure in the AO theory is replaced by that given by the 2D-SPT. In other words, the depletants collide with each other through hard-disk interaction. The coefficients $B_2/B_2^{HD}$ for the ratios $q$s are smaller than -5 when $\eta_{lip}^{res} = 0.45$. The depletion effects for the 2D-SPT reservoir are expected to be much stronger than those for the ideal gas reservoir. These results indicate that the repulsive interaction between depletants is important in these analyses.

In the present study, the depletion idea plays an important role in the ordering behaviors. Moreover, it seems that the depletion idea is also important in the association of trimer of the wild-type bRs, because the top view of a wild-type bR looks like a circular sector. When the central angle is $2\pi/3$, the trimer conformation has the smallest excluded area for the three bRs. By contrast, the top view of mutant bRs might be more circular than the circular sector, although no evidence for this was given in this work.

Finally, we discuss the validity of the assumption for the ordering structure. We assumed that the ordered state for the pure bRs was hexagonal because it seemed that the argument of the hexatic phase could be avoided in the present study. According to a simulation study for the two-component disk, the hexatic phase disappears when the number ratio of the small disk is higher than 1% [25]. The result suggests that the smaller disks stabilize the hexagonal phase. In the biomembrane, the number of smaller disks was a major component and the fragile hexatic phase could be expected to disappear. In contrast, the size ratio between disks $q$ in the simulation (about 0.714) was much larger than in this study [25]. In those smaller $q$s, such as 0.08065 and 0.16667, the small disk could locate interstitially in the ordered large disks. Therefore, further simulation studies are also needed.

In the present study, the ordering behaviors of bR was discussed. The phase diagrams for the monomers and the trimers with lipid molecules were calculated by using some simple theories with a simple 2D model. The results showed that the depletion effect was dominant for the larger hard disk ordering when the $\eta_{lip}^{res} > 0.35$. The calculated results for CCRs agreed with the experimental results, suggesting that the depletion interactions induced by the lateral translational motions of lipid molecules drive the association of membrane proteins, such as bRs.

**Acknowledgement**

The authors thank Prof. A. Yoshimori for his comments. This work was supported by Japan Society for the Promotion of Science (JSPS) KAKENHI Grants No. JP19H01863, No. JP19K03772, No. JP18H03673, No. JP18K03555, and No. JP16K05512. The computa- tions

## Supplemental Material

The second virial coefficient ($B_2$) is useful for discussing whether the particles are attractive or repulsive. The $B_2$ is defined as follows:

$$\beta p = \rho + B_2 \rho^2 + O(\rho^3). \tag{16}$$

The equation of state for an ideal gas is

$$\beta p = \rho. \tag{17}$$

Then, the $B_2$ is zero.

When the $B_2$ is positive, the effective interaction between particles is repulsive. However, when the $B_2$ is negative, the effective interaction is attractive. Here, the $B_2$ for the 2D case is calculated as follows:

$$B_2 = -\pi \int r(\exp[-\beta u(r)] - 1) dr, \tag{18}$$

where $u(r)$ is the pairwise potential and $r$ is the distance between the centers of the particles. In this study, the effective interaction $w(r)$ was substituted into the potential $u(r)$. Therefore, the $B_2$ is the effective second virial coefficient. This $B_2$ was scaled by the $B_2$ for hard disks, $B_2^{HD}$. $B_2/B_2^{HD}$ is the reduced (effective) second virial coefficient.

Two effective potentials between two large hard disks (proteins) were examined. One is the conventional AO potential [S1]. The other is a modified AO potential (MAO). In the AO theory, the depletants (lipid molecules, small disks) do not interact with each other. In other words, the depletants are ideal gas molecules. By contrast, for the MAO potential, the attraction between two large disks at the contact distance is stronger than that for the conventional AO potential, because the pressure exerted by the depletants to the large disks is calculated by using the 2D-SPT. This is because the repulsion between depletants is not ignored in the 2D-SPT and the pressure becomes larger than that of an ideal gas.

Here, two large disks (bRs) exclude the small disks. The overlap of the excluded area $\Delta V_{ex}$ depends on the distance between bRs. The effective potential $w(r)$ was calculated by using the conventional AO theory as follows:

$$w(r) = -\rho k_B T \Delta V_{ex}(r), \tag{19}$$

where $\rho$ and $r$ is the number density of lipid molecules. Here, $\rho k_B T$ is the depletant pressure $p$, and it is obtained using the equation of state for the ideal gas. Therefore, $w(r)$ is rewritten as

$$w(r) = -p \Delta V_{ex}(r). \tag{20}$$

In the case of MAO, the pressure is obtained using the equation of state for 2D-SPT as follows:

$$p = \frac{\rho k_B T}{(1 - \eta_{lip})^2}. \tag{21}$$

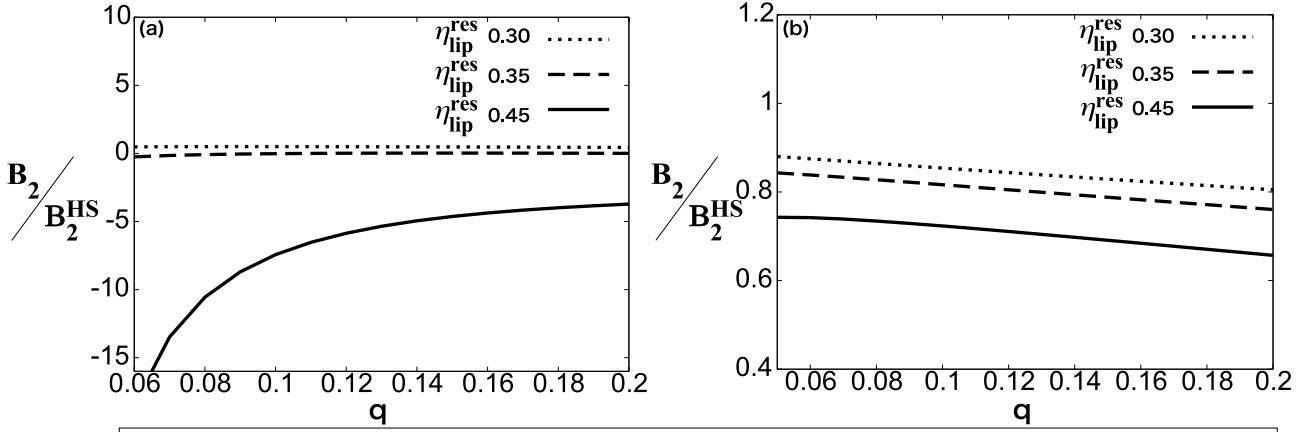

**FIG.S 1.** The reduced second virial coefficient of large disks (bRs) for MAO (a) and for AO (b) when the small disks (lipid molecules) packing fraction is 0.30 (dotted curve), 0.35 (dashed curve), and 0.45(solid curve).

This pressure $p$ is substituted to eq (20). The effective potential for MAO is shown as

$$w(r) = -\frac{\rho k_B T \Delta V_{ex}}{(1-\eta_{lip})^2}. \tag{22}$$

The reduced second virial coefficients $B_2/B_2^{HD}$ are obtained using the effective potential based on AO and MAO theories. The coefficient decreases as the packing fraction $\eta_{lip}^{res}$ becomes larger.

FIG. S1 indicates the $q$-dependence of the coefficient $B_2/B_2^{HD}$ for MAO (a) and for AO (b). The negative values appear at around $\eta_{lip}^{res} = 0.35$ in the plot for MAO (FIG.S1 (a)). In the plot for $\eta_{lip}^{res} = 0.35$, the coefficient $B_2/B_2^{HD}$ is almost zero. In addition, the negative value appears when $q$ is below 0.11. This suggests that the particles are attractive and the wide coexistence region can appear. However, the absolute value is still small when $\eta_{lip}^{res} = 0.35$.

As the packing fraction $\eta_{lip}^{res}$ becomes larger, the value $B_2/B_2^{HD}$ decreases. The $B_2/B_2^{HD}$ for MAO is smaller than 0 for any $q$ when the $\eta_{lip}^{res}$ is larger than 0.36. When the $\eta_{lip}^{res}$ is 0.45, the absolute value of $B_2/B_2^{HD}$ for MAO becomes very large because of the effective attraction. It is about -7.4 at $q = 0.1$, and we can expect the appearance of a condensation phase. In addition, FIG.S1 (a) shows that the $B_2/B_2^{HD}$ for $\eta_{lip}^{res} = 0.45$ increases monotonically as the $q$ increases. That is, the depletion attraction between bR trimers is stronger than between monomers. As we mentioned in the present article, the depletion attraction between bR trimers is stronger than that between monomers. Thus, these results for the $B_2/B_2^{HD}$ are consistent with the phase diagrams shown in the present article.

By contrast, the plots for AO do not have negative values even when $\eta_{lip}^{res} = 0.45$ (FIG.S1(b)). This means that the particles in the system are repulsive. If $\eta_{lip}^{res}$ becomes larger, the coefficient $B_2/B_2^{HD}$ becomes negative. However, the value $\eta_{lip}^{res}$ becomes unphysical. A comparison between the $B_2/B_2^{HD}$ values for MAO and AO indicates that the repulsive forces between lipid molecules are important for the association of membrane proteins.